\documentclass[fleqn,usenatbib]{mnras}

\usepackage[T1]{fontenc}
\usepackage{graphicx, amsmath, amssymb, float}
\usepackage{savesym}
\usepackage{txfonts,xcolor}
\usepackage{ae,aecompl}
\usepackage[normalem]{ulem}
\usepackage{amsmath}

\newcommand{\Msun}{{\rm M}_\odot}

\savesymbol{iint}

\renewcommand{\textbf}[1]{{\color{black} #1}}

\title[AGN vs mergers in the {\sc RomulusC} simulation]{Fountains and storms: the effects of AGN feedback and mergers on the evolution of the intracluster medium in the {\sc RomulusC} simulation
}

\author[U. Chadayammuri et al.]{
Urmila Chadayammuri,$^{1, 2}$\thanks{E-mail: urmila.chadayammuri@yale.edu}
Michael Tremmel,$^{3,4}$
Daisuke Nagai,$^{1,3,4}$
Arif Babul,$^{5}$
\newauthor 
Thomas Quinn$^{6}$
\\
$^{1}$Department of Astronomy, Yale University, New Haven, CT, 06511, USA\\
$^{2}$Chandra X-Ray Center, 60 Garden Street, Cambridge, MA, 02138, USA\\
$^{3}$Department of Physics, Yale University, New Haven, CT, 06511, USA\\
$^{4}$Yale Center for Astronomy and Astrophysics, New Haven, CT, 06511, USA\\
$^{5}$Department of Physics and Astronomy, University of Victoria, Victoria, BC V8W 2Y2, Canada\\
$^{6}$Department of Astronomy, University of Washington, Seattle, WA, 98195, USA
}

\date{Accepted XXX. Received YYY; in original form ZZZ}

\pubyear{2020}

\begin{document}
\label{firstpage}
\pagerange{\pageref{firstpage}--\pageref{lastpage}}
\maketitle

% Abstract of the paper
\begin{abstract}

{The intracluster medium (ICM) is a multi-phase environment, dynamically regulated by active galactic nuclei (AGN), the motions of cluster galaxies, and mergers with other clusters or groups. AGN provide a central heating source capable of preventing runaway cooling flows and quenching star formation, but how they achieve this  is still poorly understood. We investigate the effects of AGN feedback and mergers on the ICM using the high-resolution {\sc RomulusC} cosmological simulation of a $10^{14}$ M$_{\odot}$ galaxy cluster. We demonstrate that AGN feedback regulates and quenches star formation in the brightest cluster galaxy gently, without any explosive episodes, and co-exists with a low entropy core with sub-Gyr cooling times. In contrast, the merger disrupts the ICM structure, heating the core and cutting off the supply of low-entropy, infalling gas that until then fuelled the AGN. We find that this \textbf{removal of the low-entropy phase} correlates with the ratio t$_{\rm cool}$ / t$_{\rm ff}$ increasing above 30 in the core, matching observations that cooling gas is only found in clusters where this ratio is 5-30. \textbf{Importantly, we find that evolution in the inner entropy profile and the ratio of cooling to free-fall timescale are directly connected to the quenching of star formation in the BCG. This is in line with previous results from idealized simulations and confirmed here within a fully cosmological simulation for the first time.} 
}
\end{abstract}

\begin{keywords}
galaxies: clusters: intracluster medium --  quasars: supermassive black holes -- cosmology: large-scale structure of Universe
\end{keywords}

\section{Introduction}

Galaxy clusters are the largest and most recently formed structures in the Universe, serving as a powerful laboratory for astrophysics and cosmology. One of the outstanding questions for cluster formation is understanding the cooling and heating balance that shapes the properties and evolution of cluster cores
\citep{DeGrandi2002, McCarthy2004, McCarthy2008, Edwards2007, Bildfell2008, Sun2009, Cavagnolo2010, Voit2014, McDonald2018, McDonald2019}. The entropy of the cluster core is a simple metric that captures the net effect of heating and cooling processes. Its distribution has been observed to be slightly bimodal \citep[e.g.,][]{Cavagnolo2009, Pratt2009, Sanderson2009}, with the lower entropy clusters called cool-cores (CC) and the higher-entropy ones non-cool cores (NCC). \textbf{Specifically, the entropy profiles of CCs decline monotonically into the central few kiloparsecs, while the NCCs have a flat, high-entropy core at the center.} In CC clusters, the cores have remained relatively unchanged since z $\gtrsim$ 1.3 \citep{Santos2010, McDonald2017, Sanders2018}, suggesting that a cluster's core state is set early in its evolutionary history. 
%This is interesting because the ICM \textbf{in these clusters}, particularly its central region, is typically observed to have cooling times much shorter than a Hubble time. 
\textbf{The central regions of these CC clusters are typically observed to have cooling times much shorter than a Hubble time. The ICM is also a dynamic environment in which the gas is heated by the central AGN activities and stirred by mergers and galaxy motions at the same time.}
%- it gets heated from the center by AGN activity, gets stirred by galaxy motions within it, and undergoes mergers with other groups and clusters. It is therefore imperative to study how AGN feedback and mergers work, independently and in conjunction, to counteract or encourage cooling in the ICM and shape the evolution of cluster core.

Early simulations without feedback from active galactic nuclei (AGN) produced cooling flows, bulk flows of high-entropy gas from the outskirts towards the core, raising the core entropy over time \citep{Lewis2000, Voit2002}. Including AGN feedback suppressed these cooling flows and formed more realistic profiles, at least for the cool core category of clusters \citep{Voit2005, Sijacki2007, McCarthy2008, Guo2010}. \textbf{However, reproducing the observed distribution of cluster core entropies has been elusive.} \citet{Dubois2011}, for example, find that in a cosmological simulation with AGN feedback, including metal line cooling produces only clusters with NCC; \citet{Planelles2014} also produce a cosmological suite with AGN feedback where all the clusters resemble NCC. C-Eagle \citep{Barnes2017} and Illustris-TNG \citep{Barnes2018} include both AGN feedback and metal cooling, and do produce cool cores, although less frequently than observed. The C-Eagle study attributes the excess core entropy to heating by the AGN. \citet{Rasia2015} reproduced a realistic population of cluster cores using an improved SPH scheme along with a two-mode AGN feedback model corresponding to the quasar- and radio-modes with different efficiencies \citep{Steinborn2014}, but attribute this success primarily to improvements in the hydrodynamic solver that better capture gas mixing.

\textbf{The other major heating source are cluster mergers - the} most energetic events in the Universe since the Big Bang. With as much as $10^{64}$ ergs of initial kinetic energy, they are more than capable of significantly modifying the structure of a cluster core through shocks and mixing. Based on a joint X-ray and weak lensing analysis of observed systems, \citet{Mahdavi2013} find that cool-core clusters are more likely to be in hydrostatic equilibrium than their non-cool core counterparts, in contention with some recent simulation results \citep{Barnes2018}. \citet{Poole2006} and \citet{Hahn2017} find that low angular momentum mergers do not always disrupt CCs and even when they do, the central cooling time often does not increase to over $0.1t_H$. Overall, the emerging picture is that while mergers can heat cluster cores, the parameters of a merger, such as merger mass ratios and impact parameters, likely play a critical role. Even so, studies have found that in the presence of radiative cooling, the net heating lasts only 3-4 Gyrs, after which the core returns to its pre-merger entropy.
%\citep{Poole2008, Burns2008}.  

Recent observations of H-$\alpha$, infrared and molecular line emission have emphasised the multi-phase nature of the ICM, and the role of cold gas in mediating the interaction between AGN and the ICM \citep{Salome2006, Peterson2006, Risaliti2010, Grier2012, Tremblay2016}. It should be noted that the key phases of the ICM are different from that of the interstellar medium, with the cool phase usually referring to gas at or below the H-$\alpha$ emitting threshold of $10^4$K \citep[e.g.,][]{Gaspari2012, McCourt2012}. 
%\citet{Donahue2004} notes that even though this is hot by some standards, it is orders of magnitude below the cluster virial temperature, so the gas must have either avoided gravitational heating or experienced significant cooling. 
This H$-\alpha$ emission has only been observed in clusters with central entropies below 30 keV cm$^2$ and correspondingly low central cooling times \citep{Crawford2005, Salome2006, Cavagnolo2008, Mittal2009, Hlavacek-Larrondo2012, Voit2015}. Clusters with higher core entropies, on the other hand, had longer cooling times that flattened in the central regions and no detectable cold gas, and were much less likely to host an AGN, indicating that the AGN-ICM regulation is disrupted, or non-existent, in these systems.

Since the cold gas has a very low volume filling factor, capturing the formation of the multi-phase ICM requires resolution that has so far remained in the domain of idealised simulations. These have shown that the warm-hot ICM in the cluster core is in approximate global thermal balance, and becomes susceptible to local thermal instabilities when the ratio of the cooling time ($t_{\rm cool}\equiv E_{\rm therm}/[n^2\Lambda(T, Z)]$, where $\Lambda$ is the gas temperature and metallicity-dependant radiative cooling function) and the gravitational free-fall time ($t_{\rm ff}\equiv \sqrt{2r/g}$, where $g=GM(<r)/r^2$), drops below a threshold typically in the range 5-30 \citep[e.g.,][]{McCourt2012,Sharma2012,Prasad2015,Voit2017}. At this point, cold dense clouds start to condense out of the warm-hot ICM, fall towards the cluster centre, and eventually accrete onto the central black hole \citep[e.g.,][]{Gaspari2013,Voit2015,Li2015, Lakhchaura2018}. In this model, black hole accretion is set by the rate at which gas is able to `precipitate' out of the hot/warm ICM, which in turn feeds the central SMBH; the resulting feedback regulates the cooling in the cluster core \citep{Gaspari2017a,Prasad2017,Prasad2018}.

Idealised simulations allow controlled experiments and high resolution, but crucially lack cosmological context. The amount of cold gas fueling the AGN activity is strongly boosted by the presence of turbulence \citep{Gaspari2013, Prasad2017}, \textbf{but the bulk and turbulent motions generated by the cosmic accretion process are either left out or must be introduced by hand in an idealized way.}
%\textbf{idealised simulations have to stir in explicitly}. 
In cosmological environments, bulk and turbulent gas motions arise naturally from mergers and interactions of galaxies and galaxy groups within the cluster \citep{Lau2009, Vazza2011, Nagai2013, Nelson2014} and penetrating streams from the cosmic web \citep{Zinger2016}. It is therefore useful to study the effects of AGN feedback and mergers on the evolution of cluster cores by using a fully cosmological simulation capable of capturing the roles of AGN feedback in a realistic cosmological context \textbf{while resolving rich structure within the cool-warm ICM down to 100s pc scales}.
%\textbf{resolving multiple phases of the ICM}. \textbf{I think this is actually the opposite of what the second referee wanted. They said to remove references to resolving the multiphase ICM. I think this is too harsh and would prefer a more precise statement saying that we can resolve the multiphase strucutre of the ICM down to $\sim 100 pc$ scales which I think is fairly accurate. In Butsky et al 2019 paper we use the phrase multiphase cool-warm gas to specify $10^{4}-10^{7}$K gas}.

In this paper, we use the high-resolution, cosmological, hydrodynamic, zoom-in galaxy cluster simulation {\sc RomulusC} \citep{Tremmel2019} to study in detail the effects of AGN feedback and mergers on the thermodynamics of the ICM. {\sc RomulusC} simulates a low mass galaxy cluster with $M_{200} = 1.4 \times 10^{14} M_{\odot}$ \textbf{ and emission-weighted $kT_{500} = 1.4$keV} at $z=0$. With spatial and mass resolution of 250 pc and $10^5 M_{\odot}$, respectively, {\sc RomulusC} approaches resolutions previously only attainable in idealized simulations of isolated clusters. At this resolution, {\sc RomulusC} produces rich structures of gas with \textbf{$T\gtrsim 10^4$~K} \citep{Butsky2019}. \textbf{For the purposes of this paper, we define "cooler gas" as that with $T < 10^5K$.}
%, which is two orders of magnitude cooler than the volume-filling, X-ray emitting gas.} 
This resolution also makes it one of the highest resolution cosmological simulations of a halo of this mass to date, comparable only to the most massive halo from the TNG50 simulation \citep{Pillepich2018, Nelson2019}. As discussed in \citet{Tremmel2019}, the ICM core in {\sc RomulusC} remains at low entropy with sub-Gyr cooling times until the onset of a 1:8 mass ratio merger at z$\sim$0.14. Prior to this merger, star formation quenches due to AGN feedback. This simulation therefore offers a unique case study of the effects of AGN feedback and mergers on the ICM.

We describe the {\sc RomulusC} simulation and its key baryonic physics prescriptions in $\S$\ref{sec:sim}. $\S$\ref{sec:results} presents the evolution of thermodynamic profiles, cooling time and precipitation rate, connects this to the behaviour of cooling, multi-phase gas at all radii. We discuss our results in light of previous work and important caveats in $\S$\ref{sec:discussion}. Conclusions are summarized in $\S$\ref{sec:conclusions}.

\section{The {\sc RomulusC} Simulation}
\label{sec:sim}
{\sc RomulusC} \citep{Tremmel2019} is a zoom-in cosmological simulation of a galaxy cluster with initial conditions extracted from a 50 Mpc-per-side dark matter-only simulation using the standard volume re-normalization technique of \citet{Katz1993} and re-run with higher resolution and full hydrodynamic treatment using the Tree+SPH code, {\sc ChaNGa} \citep{Menon2015}. The {\sc Romulus} simulations use a spline kernel force softening of 350 pc (corresponding to a Plummer softening of 250 pc) and a  minimum SPH smoothing kernel of 70pc. With dark matter and gas mass resolutions of $3.4\times10^5\Msun$ and $2.12\times 10^5\Msun$ respectively, {\sc RomulusC} resolves substructure down to $3\times10^9\Msun$ with over $10^4$ particles, naturally including interactions between the ISM of in-falling galaxies with the ICM of the cluster environment. The cluster reaches M$_{200}$ of $1.15\times10^{14}M_\odot$ by $z=0$. {\sc ChaNGa} uses prescriptions for gas cooling, star formation and feedback, and UV background from {\sc Gasoline} \citep{Governato2010}, turbulent dissipation \citep{Wadsley2017} and introduces a new model for supermassive black hole (SMBH) accretion, feedback, and dynamical evolution. All the free parameters for sub-grid models related to SMBH or stellar physics were tuned to reproduce a series of scaling relations for galaxies of Milky Way mass or smaller \citep{Tremmel2017}, so that results on the cluster scale are purely predictions of the model. The simulation assumes Planck cosmology: $\Omega_m$ = 0.3086, $\Omega_\Lambda$ = 0.6914, h= 0.67, $\sigma_8$= 0.82 \citep{Planck2015}. 

\subsection{Star formation and gas physics}
\label{sec:sf_gas}
Star formation occurs in gas particles with density greater than 0.2$m_p$ $cm^{-3}$ and temperature below $10^4$K. A star particle of mass $m_*$ forms out of a gas particle of mass m$_{gas}$ with the probability:
\begin{align}
p = \frac{m_{\rm gas}}{m_*}\left(1 - \exp(-c_*\Delta t/t_{\rm dyn})\right),
\end{align} 
where the local dynamical time is $t_{\rm dyn}= \sqrt{3\pi/(32G\rho)}$, the characteristic timescale is taken as $\Delta t = 10^6$yr for star formation, and the star formation efficiency $c_*$ = 0.15. The latter is a free parameter tuned in \citet{Tremmel2017} to reproduce a series of scaling relations for MW-sized and dwarf galaxies. Star particles are assumed to follow a Kroupa IMF \citep{Kroupa2001} to compute metal enrichment and supernova rates. Supernova feedback is implemented as a thermal injection with 75$\%$ coupling efficiency accompanied by a cooling shutoff in the surrounding gas, following the `Blastwave' formalism from \citet{Stinson2006}. 

ChaNGa includes an updated SPH implementation that uses a geometric mean density in the SPH force expression, allowing for the accurate simulation of shearing flows with Kelvin-Helmholtz instabilities \citep{Ritchie2001, Governato2015, Menon2015}. The most recent update to the hydrodynamic solver is the improved implementation of turbulent diffusion \citep{Wadsley2017}, which does not occur on sub-resolution scales in traditional SPH codes (see \citealt{Rennehan2019} and references therein). In the presence of gravity, where buoyancy and conduction lead to the separation of low and high-entropy gas, traditional SPH codes produce unphysically low-entropy cores \citep{Wadsley2008,Mitchell2009} and fail to distribute metals over large radii \citep{Shen2010,Rennehan2019}. The combination of a gradient-based shock detector,time-dependent  artificial  viscosity,  and  an  on-the-fly  time-step adjustment system allows for a more realistic treatment of both weak and strong shocks \citep{Wadsley2017}. The CHaNGa solver is thus better numerically equipped to study the thermodynamics of turbulent environments, such as the ICM with frequent merger activity and AGN outflows.

As discussed in \citet{Tremmel2019}, an important limitation of {\sc RomulusC} is the lack of high temperature metal line cooling, a major coolant for gas in high mass halos such as the one we examine here. While the lack of metal line cooling certainly can affect the accretion of gas onto central galaxies, feedback processes, particularly from AGN, are also important \citep{vdVoort2011a}. This choice was made because the resolution of {\sc RomulusC}, while unprecedented, is insufficient to resolve molecular hydrogen. \citet{christensen14b} find that the inclusion of metal line cooling without a star formation prescription that accounts for molecular hydrogen results in overcooling. One possible solution to the overcooling problem would be to boost feedback efficiency \citep[e.g.,][]{shen12,dallavecchia12,Schaye2015,sokolowska16,sokolowska18}, but this will not necessarily provide a realistic ISM or CGM/ICM \citep{sokolowska16}. Another potential solution would be to only include metal line cooling in diffuse gas, but determining an arbitrary threshold below which unresolved multiphase structure exists is difficult and may have unforeseen affects on galaxy evolution. We therefore opt to not include it here. We discuss further in \S4.4 how this lack of metal cooling may influence our conclusions.

\begin{figure*}
    \centering
    \includegraphics[width=\textwidth]{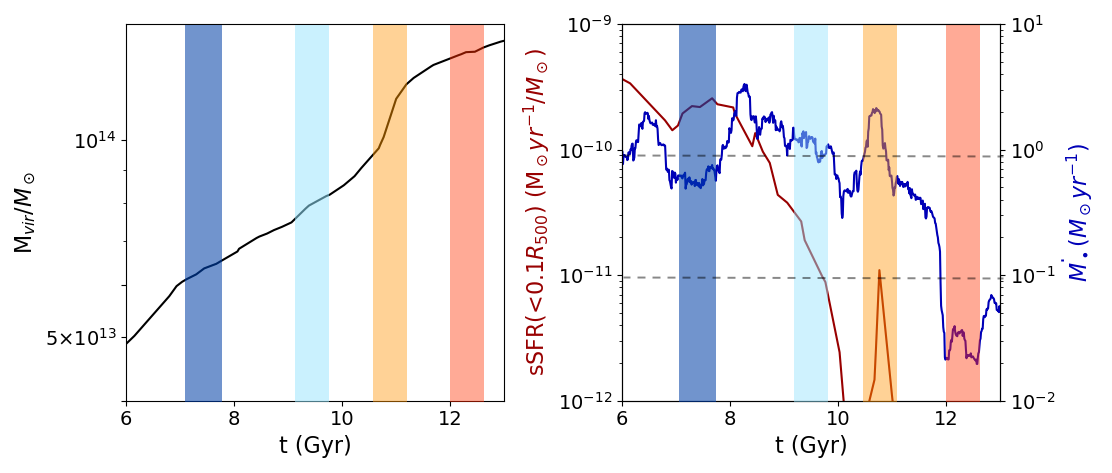}
    \caption{\textit{Left panel:} Evolution of the virial mass of the {\sc RomulusC} cluster over time. \textit{Right panel:}  Black hole accretion rate (blue) and specific star formation rate in the central 10kpc (red), both smoothed over 250Myr, over time in {\sc RomulusC}. In this work, we specifically investigate the four key epochs indicated in the shaded color: before (purple) and after (blue) quenching, between quenching and beginning of the merger (orange), and during the merger (red). \textbf{The horizontal dashed lines represent accretion rates corresponding to feedback power $10^{44}$ and $10^{45}$ ergs/s, assuming a radiative efficiency of 10\%}.}
    \label{fig:bcg_history}
\end{figure*}

\subsection{The Romulus SMBH Model}
The model for SMBHs and the AGN they generate in {\sc RomulusC} differs from standard implementations in several key ways. First, black holes are seeded based on gas particle properties, rather than the host halo mass, as is common in the literature. Second, accretion onto SMBHs accounts for the presence of rotationally supported gas nearby the black hole. Third, the feedback model avoids prescribing a mass or accretion rate dependence to the feedback mode and instead only includes a simple model for thermal feedback which, thanks to the resolution and implementation of the model, naturally results in large-scale outflows that can quench star formation in massive galaxies \citep{Pontzen2017, Tremmel2019}. The SMBH model is described in further detail in \citet{Tremmel2017}, so we only provide a brief overview here to highlight the aspects relevant to the AGN-merger connection and impacts on core gas properties.

\subsubsection{Black hole seeding}
Black holes are initially seeded in the simulation from a sub-set of potentially star-forming gas particles that (i) have densities that exceed 15 times the threshold for star formation, (ii) have very low metallicity (Z $< 3\times 10^{-4}$), and (iii) have a temperature just below the limit of atomic cooling, 9500-10$^4$K. This physically identifies dense, quickly collapsing regions that are most likely able to grow a 10$^6$ M$\odot$ black holes in the early Universe. Most other cosmological simulations posit instead that every halo will contain a SMBH by the time its mass hits a certain threshold, usually 10$^{10-12 }M_\odot$ \citep[e.g.,][]{Sijacki2005, Steinborn2014, weinberger_2016, Stevens2017}. In comparison, SMBHs in {\sc Romulus} are seeded at higher redshift and in lower mass halos (10$^8-10^9M_\odot$) in {\sc RomulusC}. AGN feedback in {\sc RomulusC} halos is allowed to begin much earlier ($z\gtrsim5$) and in lower mass halos compared to other simulations of similar scale. \citet{Ricarte2019} find that in {\sc Romulus} SMBH growth (and therefore feedback) traces the star formation of galaxies even at early times and in low mass halos. These results appear to be in contrast with other theoretical results showing a suppression of SMBH growth in low mass galaxies due to supernovae feedback \citep{Dubois2015, Habouzit2017, Bower2017, AnglesAlcazar2017}. 

We note that a lack of sustained SMBH growth does not itself prove that feedback from SMBHs is unimportant. \citet{Sharma2019} show that AGN feedback can have an effect on the evolution of low mass galaxies, which further highlights the importance of modeling these feedback processes in the early Universe. The high density environment of clusters may also result in unique episodes of AGN feedback as galaxies experience ram pressure \citep{Ricarte2020}.

\subsubsection{Black hole accretion and feedback}

 Accretion is a slightly modified version of the \citet{Bondi1952} model:
\begin{align}
    \dot{M_\bullet}=\alpha \times\left\{
                \begin{array}{ll}
                  \frac{\pi(GM_\bullet)^2\rho}{(v^2_{bulk}+c_s^2)^{3/2}} &\textrm{ if } v_{\rm bulk} > v_\theta\\\\
                  \frac{\pi(GM_\bullet)^2\rho c_s}{(v^2_{bulk}+c_s^2)^{2}} &\textrm{ if } v_{\rm bulk} < v_\theta
                \end{array}
              \right.
 \end{align}
 
 \begin{align}
    \alpha =\left\{
                \begin{array}{ll}
                  \left(\frac{n}{n_{th,*}}\right)^\beta & \textrm{ if } n\geq n_{th, *}\\\\
                  1 &\textrm{ if } n < n_{th,*}
                \end{array}
              \right.
 \end{align}
where $\rho$, $c_s$ and $v_{\rm bulk}$ are the density, sound speed and bulk velocity of the gas particles neighboring the black hole, and $n_{th, *}$ the critical density for star formation. The density dependent boost factor $\alpha$, identical to that of \citet{Booth2009}, acknowledges that gas with densities $n$ higher than the threshold for star formation $n_{th,*}$ will have unresolved, multi-phase structure. Calibrating against a series of observed black hole scaling relations, the best-fit value for $\beta$ was 2 \citep{Tremmel2017}. Importantly, this definition for the boost factor only takes effect for high density gas. Even at this high resolution, structures below $10^{4}$ K and at densities higher than $\sim0.2$ m$_p$/cm$^{-3}$ are unresolved due to the limited mass ($\sim10^{5}$ M$_{\odot}$) and spatial (250 pc) resolution. Gas at such high densities likely has considerable, unresolved multiphase structure that will result in under-predicted accretion rates if unaccounted for.

A key difference from the Bondi model is the introduction of $v_\theta$, the rotational velocity of the gas around the black hole. This explicitly accounts for rotational support against free fall into the black hole. $v_\theta$ is computed as the velocity a parcel of gas would have at one softening length $\epsilon_g$ away from the black hole, if angular momentum from larger scales were conserved. This introduces a coupling between feedback and gas kinematics. When there is little to no coherent rotation measured at these scales, the model reduces to the Bondi rate boosted by the factor from \citet{Booth2009}.

Energy is continuously fed back into the surrounding gas every black hole timestep as a thermal injection. Each time a particle receives feedback energy from a black hole, its cooling is shut off for a duration equal to one black hole time step, similar to the blastwave model of \citet{Stinson2006}. Black holes and the gas particles around them are placed on the smallest global timestep, typically $\sim10^4$ yrs, so that the cooling shutoffs are short compared to the local free fall time (3-5Myr) so that thermal properties are more continuously sampled when calculating accretion and feedback. The cooling shutoff aims to address the artificial overcooling resulting from insufficient resolution, and has been shown to produce better convergence in gas properties than other versions of thermal feedback and kinetic/momentum-driven feedback \citep{Smith2017}. In this context, the cooling shutoff, in combination with the very short SMBH timesteps, captures the fact that energy from growing SMBHs is transferred continuously rather than instantaneously. In this implementation, gas properties are able to respond to the transfer of energy before immediately -- and unrealistically -- radiating that energy away. The result is that this simple feedback prescription is able to drive powerful outflows out to 10s of kpc \citep{Pontzen2017, Tremmel2019} in a way that is predictive, rather than a choice explicitly made in the sub-grid model.

The feedback energy is proportional to the accretion rate, 
\begin{align}
E_{FB} = E_r E_f \dot{M}_\bullet c^2 dt,
\end{align}
where the radiative efficiency $E_r$ is assumed to be 0.1 and the best fit value for $E_f$ was found to be 0.2 \citep{Tremmel2017}. The numerical resolution of {\sc RomulusC} allows for the injection of thermal energy from SMBH feedback to occur on scales of $\sim$100pc. This is observed to drive outflows on scales of 10s-100s of kpc \citep{Tremmel2019, Pontzen2017} with neither explicit large-scale implementations nor momentum injection. This feedback prescription, unlike recent simulations in the literature \citep{Sijacki2005, Dubois2013, Sijacki2015, Weinberger2017, Oppenheimer2018}, does not distinguish between quasar and radio mode at different accretion rates. These simulations distinguish between an inefficient radiative/quasar mode and a more efficient kinetic/radio mode occuring typically for more massive black holes with lower eddington ratios. We choose to keep the feedback prescription simple, since the microphysics of AGN feedback are still poorly understood and are sensitive to parameters like black hole spin and magnetic fields that are beyond current detection methods. Besides producing large-scale outflows, our constant efficiency thermal feedback with a cooling shutoff reproduces scaling relations for galaxies at a range of masses \citep{Tremmel2017}. We refer the reader to $\S$2.2 of \citet{Tremmel2019} for a more detailed discussion.

During the epochs considered in this simulation, the mass of the SMBH in {\sc RomulusC} doubles from $7\times10^9$ to 1.4$\times 10^{10} M_\odot$. The AGN feedback power fluctuates rapidly around a mean total luminosity (assuming 10\% radiative efficiency and 2\% feedback efficiency) of $\sim10^{44}$~erg/s while the cluster is isolated and falls to $\sim 2\times10^{42}$~erg/s after the merger; and only 2\% of the radiation energy couples to heat the gas.

\section{Results}
\label{sec:results}

\begin{figure*}
\begin{center}
\includegraphics[width=\textwidth]{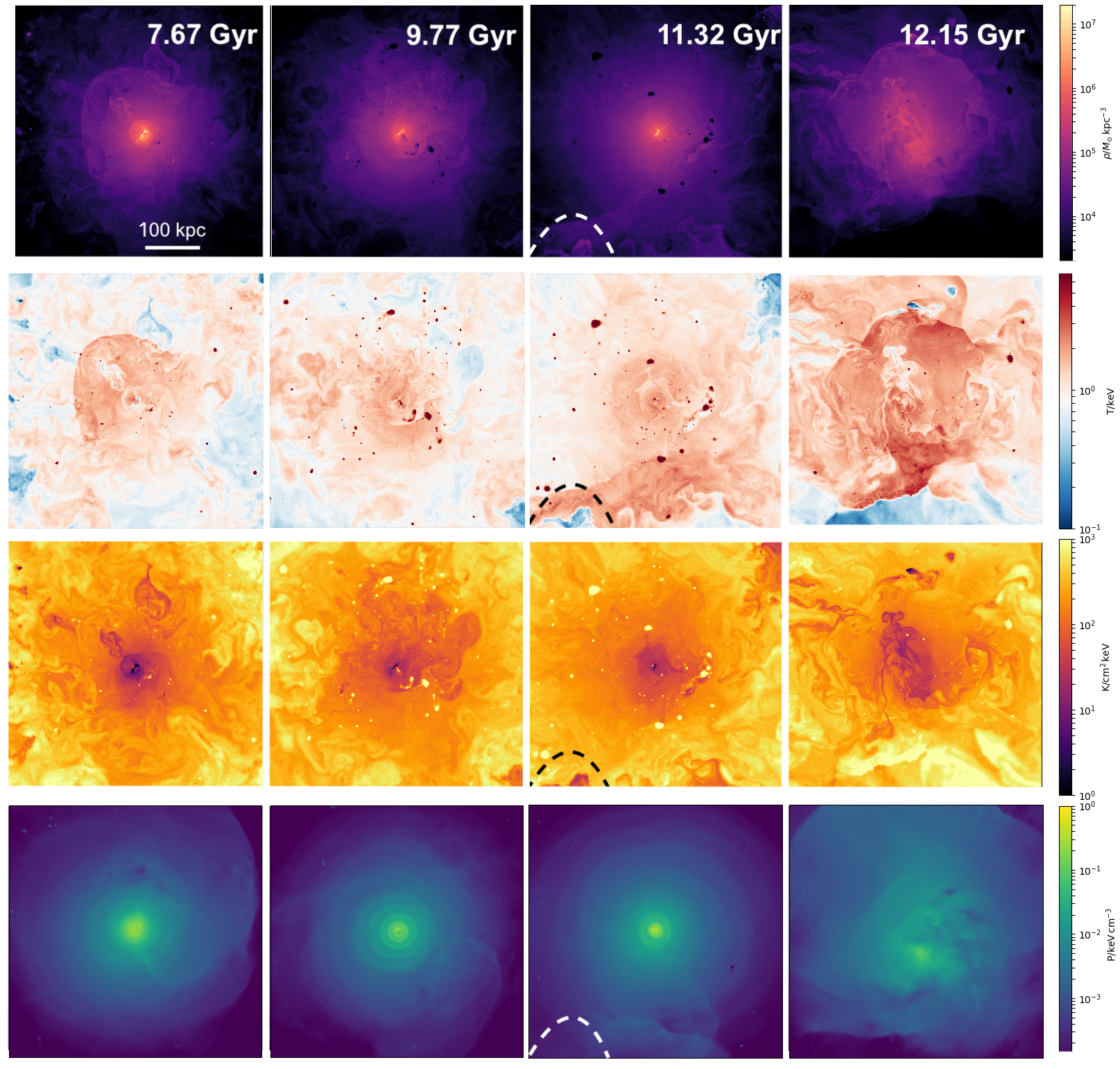}
\caption{From top to bottom: ICM column density, mass-weighted temperature, mass-weighted entropy, and volume-weighted pressure maps through a slice with 10kpc thickness of {\sc RomulusC} at the four key epochs analyzed in this work: low ({\it first column}) and high ({\it second column}) AGN activity before merger, once the central galaxy has been quenched ({\it third column}) and at the beginning of the merger ({\it last column}). Each map is 500~kpc a side, projected over 10~kpc along the line of sight. The in-falling substructure, whose northern outskirts are seen to the bottom of the images in the third column, disrupts the ICM structure shortly after it begins to interact with the cluster core ($t=12.15$ Gyr, fourth column). The core remains in this disrupted state, characterized by a marked decrease in cooling and AGN activity, at least until the end of the simulation 1.65 Gyr later. }

\label{fig:mergermaps}
\end{center}
\end{figure*}
\subsection{The four epochs of cluster dynamics }

We consider the cluster at four different epochs shown in Figure~\ref{fig:bcg_history}, chosen to represent four important stages of evolution for the brightest cluster galaxy (BCG).
%and \textbf{the cooler gas with $10^4K < T < 10^5$K in the ICM}. 
The duration of each epoch is $\approx 0.5$Gyr, about twice the dynamical time in the core, so that stochasticity in the AGN activity does not dominate our analysis. The first period (7.3-7.7 Gyr, dark blue) captures the cluster when it is isolated and star-forming. The next epoch (9.2-9.7 Gyr, light blue) is characterized by the BCG actively quenching, coincident with a period of enhanced AGN activity and outflows as discussed in \citet{Tremmel2019}. The third epoch (10.8-11.3 Gyr, orange) occurs after the BCG is fully quenched but the cluster is still isolated, and the fourth (12-12.5 Gyr, red) takes place during the first pericenter passage of a merger with a galaxy group, which was approximately one-eighth the mass of the main halo at in-fall. We keep these colors consistent throughout the following analyses.

The left panel shows the evolution of the virial mass of {\sc RomulusC} over cosmic time. The epochs that we study begin when the cluster progenitor is $\sim6\times10^{13}$ M$_{\odot}$ in mass. On the right panel, we plot the specific star formation rate (red) and black hole accretion rate (blue) as a function of time. \textbf{The horizontal dashed lines indicate the accretion rates that correspond to the AGN feedback rates of $10^{44}$ and $10^{45}$~erg/s.} The light blue epoch witnesses a decline in specific star formation rate (sSFR) of two orders of magnitude, and the orange epoch encompasses a slight resurgence before the quenching is complete. The red epoch witnesses the quenching of black hole activity. The $r_{500}$ at each of these epochs is 287, 388, 481 and 601 kpc, respectively. The mass of the black hole increases from 8.5 $\times 10^9 M_\odot$ to $1.3 \times 10^{10}M_\odot$, with most of the growth happening before the merger begins at 11.65 Gyr. \textbf{The emission-weighted $kT_{500}$ ranges from 1.2 to 2.0~keV over this period, before reverting to 1.4 keV at z = 0.}

\begin{figure*}
    \centering
    \includegraphics[width=\textwidth]{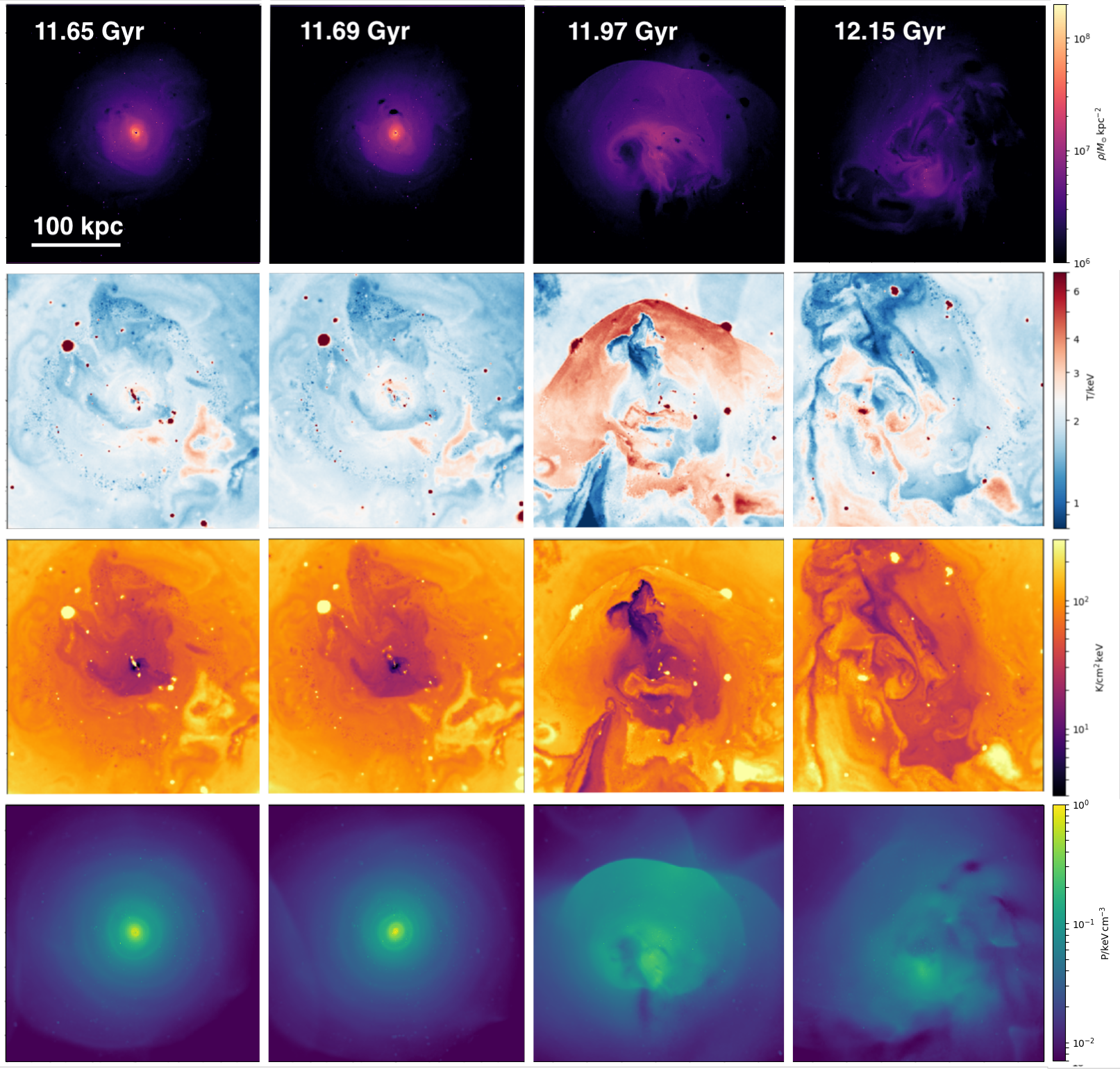}
    \caption{Maps of column density, mass-weighted temperature, mass-weighted entropy and volume-weighted pressure through a slice with 10~kpc thickness from the halo center immediately around the disruptive merger event. The shown region is 300kpc in size. The substructure is outside this region in the first two snapshots, punches through the core of the main halo at 11.97 Gyr, creating a high pressure bubble that expands and heats the ICM by 12.15 Gyr. Throughout this time period, AGN activity is 1-2 orders of magnitude lower than when the cluster was isolated, so that it cannot cause the heating. }
    \label{fig:shock}
\end{figure*}

Figure~\ref{fig:mergermaps} shows the column density, mass-weighted temperature, mass-weighted pseudo-entropy $K \equiv k_BT\times n_e^{-\frac{2}{3}}$, and volume-weighted pressure $P \equiv k_BT\times n_g$ along the line of sight through a slice with 10 kpc in thickness during each of these epochs; each image represents a 500~kpc-per-side view of the cluster. The weighting for each quantity is chosen to most closely resemble the observable analogue.

Despite star formation in the BCG declining by over two orders of magnitude, the structure of the ICM remains relatively unchanged until the right-most panel depicting the time of the merger event. Figure~\ref{fig:shock} zooms in on the four snapshots immediately surrounding the first pericenter passage of the merging substructure (now with 300 kpc-per-side). The top three maps clearly show bulk motions of gas, which takes high density, low temperature gas from the core and mixes it with warmer gas at larger radii. The temperature map shows that gas is also heated during the pericenter passage. A shock wave is first seen at 11.97Gyr, and expands and diffuses within a radius of 150kpc by the next snapshot 0.2Gyr later. The disruption of the low entropy core thus occurs not only from removing cool, dense gas from the core and mixing it with less dense, hotter gas from further out, but also by shock heating of the central gas by the merging substructure. This heating channel is also found to be significant in idealised simulations \citep{McCarthy2004,Poole2006,Poole2008}.

\begin{figure*}
\begin{center}
\includegraphics[width=0.475\textwidth]{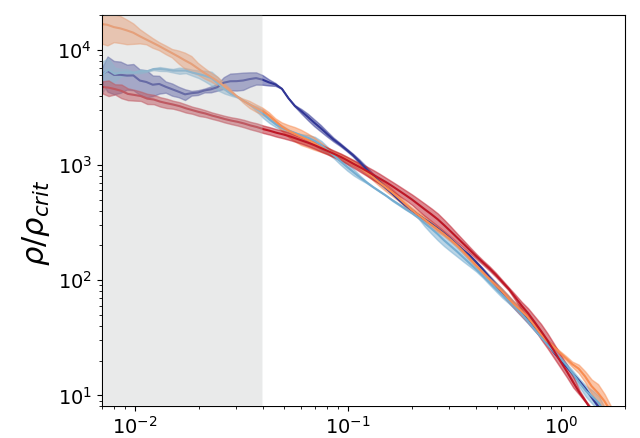}
\includegraphics[width=0.475\textwidth]{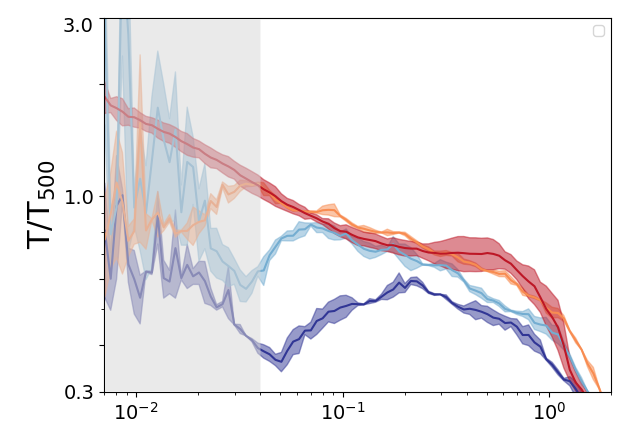}
\includegraphics[width=0.48\textwidth]{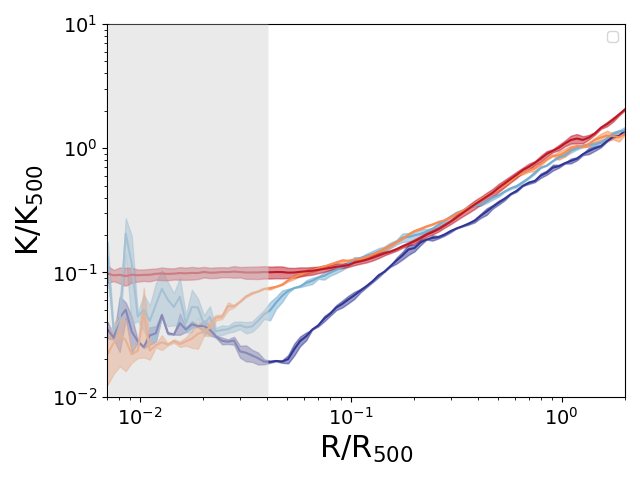}
\includegraphics[width=0.48\textwidth]{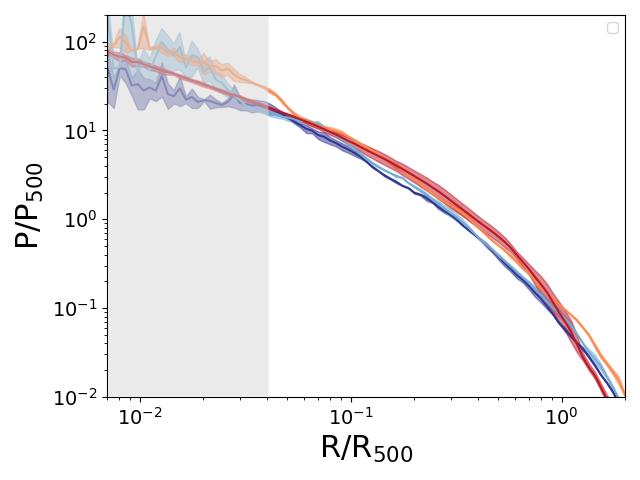}
\caption{Evolution of 3D gas mass density ({\it top-left}), mass-weighted temperature ({\it top-right}) and entropy ({\it bottom-left}), and volume-weighted thermal pressure ({\it bottom-right}) derived from the volume averaged electron number density and mass-weighted temperature in the ICM. Each color represents each of the four key epochs: before (purple) and after (blue) quenching, between quenching and beginning of the merger (orange), and during the merger (red). All quantities are normalised by the value at R$_{500}$, with the self-similar values for $K_{500}$ and therefore $T_{500}$ derived in \citet{McCarthy2008}. Since the quantities fluctuate within each 0.5 Gyr epoch, the solid line shows the mean value and the shaded region shows the 1-$\sigma$ variation. The thermal pressure remains roughly constant throughout, validating its use as a mass proxy regardless of the dynamical state of the cluster. } 
\label{fig:ICMprofiles}
\end{center}
\end{figure*}

Figure~\ref{fig:ICMprofiles} shows profiles for the volume-averaged density, mass-weighted temperature, entropy, and pressure during each of the four epochs discussed above. Similar to observations, entropy and pressure profiles were derived from the mass-weighted temperature and volume-averaged electron number density within a given radial bin. All radial profiles are computed for gas hotter than $10^6$K, corresponding to the temperatures detectable by Chandra ($>0.1$ keV). 
We shade the central 0.04R$_{500}$, the largest size of the BCG, \textbf{ which conservatively denotes} the area experiencing noisy behavior due to stellar and SMBH feedback processes associated with the BCG. This was estimated by examining the gas content of the simulation (see, e.g., Figure 10 from \citet{Tremmel2019}) for the region dominated by the cold ISM disk. This region encompasses the vast majority of star formation prior to quenching. Within this region there are strong fluctuations due to stellar and black hole feedback processes. These fluctuations are absent in the epoch of merger (indicated in red), where the cold gas disk has been destroyed by the merger. All the profiles are self-similar, \textbf{i.e., described by gravitational collapse alone \citep{Kaiser1986},} 
outside $0.1R_{500}$ ($\sim$60 kpc at z = 0) for all but the earliest epoch, where the cluster seems to still be forming; there, the temperature and entropy profiles are self-similar outside 0.2$R_{500}$. Secular processes in the ICM do increase the entropy in $r=(0.1-0.3)R_{500}$ over time, but only after the merger is the entropy significantly elevated and flattened within the central $r\lesssim 0.1R_{500}$. \textbf{To account for the change in background density and temperature expected in the $\sim5$~Gyr considered here, the density is normalised by the critical density of the Universe at the corresponding redshift; the appropriate normalisation $K_{500}$ was derived in \citet{McCarthy2008}, and $T_{500}$ and $P_{500}$ follow simply thereafter.}

When the cluster core is isolated and AGN activity is low (dark blue), the entropy profile declines all the way to the central $0.05R_{500}$. A  prolonged episode of AGN feedback quenches star formation in the BCG \citep{Tremmel2019} while slightly increasing entropy at intermediate scales at $r=(0.03-0.15)$R$_{500}$ (light blue).The profile maintains this form as the BCG gets completely quenched and the AGN activity spikes yet higher in the orange band. \textbf{Importantly, the gradient of the profile becomes more positive in the light blue epoch compared to the dark blue, and in the orange compared to the light blue. In this latter, fully quenched epoch, the entropy gradient is positive even in the central 10 kpc. Idealised studies \citep[e.g.,][]{McCourt2012, Voit2018} have shown that such a positive entropy gradient, along with $t_{\rm cool}>> t_{\rm ff}$, suppresses the condensation of low-entropy gas out of a warm-hot ICM, and we now see the same effect in our cosmological simulation.} \textbf{It is important to note that this inner region of the simulation is dominated by the galaxy and is therefore strongly affected by various feedback processes.} Only after the merger (red) does the shape of the entropy profile change; it is now flat within 0.1$R_{500}$, or 60 kpc at that time.

\begin{figure*}
\begin{centering}
\includegraphics[width=\textwidth]{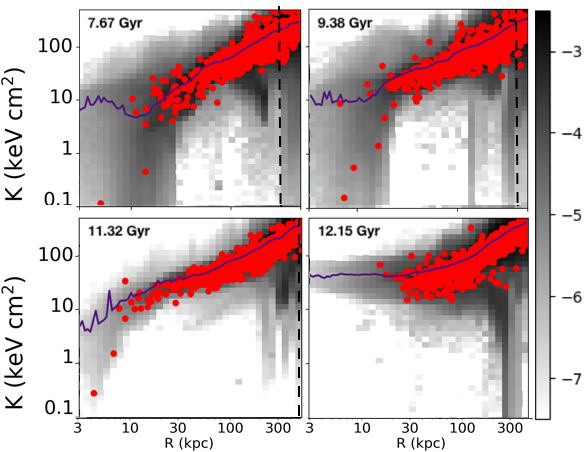}
 \caption{The entropy distribution of all the gas particles within 500kpc of the halo centre. The purple line shows the mass-weighted average of the hot ($T>10^6$K) X-ray emitting gas. Red points \textbf{represent a subset of the} particles that received \textbf{feedback energy} from the AGN any time after $t=6$Gyr. The colour map is the log of the probability distribution of particles at a given point in the grid. The black vertical dotted lines show $R_{500}$ at the corresponding snapshot; it is outside the depicted region in the last panel. The orange dashed lines indicate the shaded region in the profile plots. AGN heating clearly removes gas from the cluster centre, since the red particles rapidly settle at $r>30$kpc where their entropy is close to the average profile. However, the entropy in the inner $\sim$30kpc remains low until the merger, with plenty of gas significantly lower in entropy than the average. The merger diminishes the amount of \textbf{low-entropy} gas at all radii, and elevates the core entropy within 1 Gyr of the subshalo entering the virial radius of the main halo.}
\label{fig:multiphase}
\end{centering}
\end{figure*}

In the lower right panel of Figure.~\ref{fig:ICMprofiles}, we see that neither the merger nor AGN activity affects the thermal pressure profiles. This is in agreement with observational and theoretical evidence that cluster pressure profiles exhibit a remarkable level of self-similarity \citep{Nagai2007,Arnaud2010} and that its integrated counterpart, the core-corrected Compton Y, ought to be robust mass proxy \citep{Nagai2006,Kravtsov2006,Poole2007} even in presence of AGN feedback and mergers. 

\subsection{AGN feedback and mergers as regulators of gas cooling}

\begin{figure}
\includegraphics[width=0.475\textwidth]{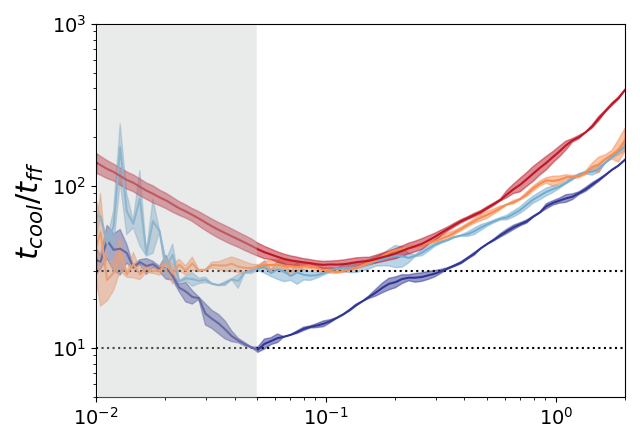}
\includegraphics[width=0.475\textwidth]{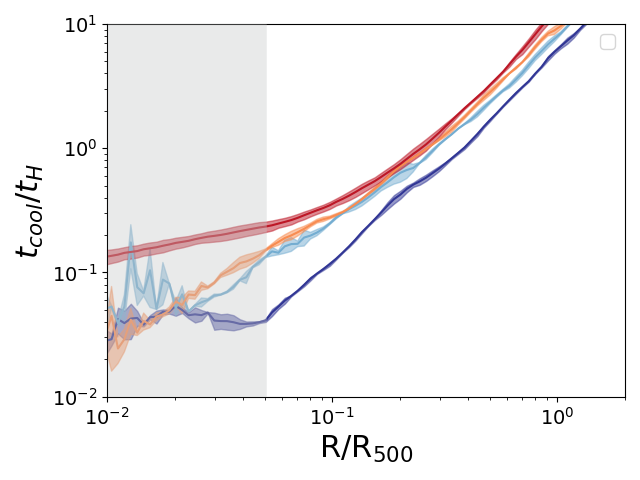}
\caption{{\it Top panel:} The ratio of cooling time to free-fall time as a function of cluster-centric radius. As in Figure~\ref{fig:ICMprofiles}, the shaded region indicates the extent of the BCG at the earliest epoch; at later times, it is a slightly smaller fraction of $R_{500}$. At early times when the BCG is actively forming stars, the ratio is as low as 10. During the period when AGN activity is quenching star formation and decreasing the amount of \textbf{cooler} gas in the center of the cluster, the ratio hovers around 30. The dashed lines represent the approximate unstable region predicted by theory and supported by observations. AGN feedback is able to increase this ratio to avoid runaway cooling and star formation. {\it Bottom panel:} The cooling time as a function of cluster-centric radius. The evolution seen in the top panel is due to an increase in the cooling time, particularly within $\sim0.1R_{500}$ as well as a steepening of the cooling time evolution with cluster-centric radius. Even as star formation quenches in the BCG, the cooling times are significantly below a Hubble time. While the cooling times are increased by a factor of several during the merger at the end of the simulation, they are still as low as 10\% the Hubble time. The on-going merger is a likely source for continued heating at $z=0$, but this suggests that the low entropy, actively cooling core is likely to reform within a few Gyr following the merger in absence any continued heating source.}
\label{fig:tcool_tff}
\end{figure}

Analytical work and idealised simulations have also shown that the infall of cooling gas depends crucially on the shape of the entropy profile. Small fluctuations in entropy create many small pockets of gas that are cooler than their surroundings. These parcels move radially inward, until they reach a region where they match the local entropy \citep{Voit2018}. If the entropy profile is flat within a certain radius, gas that is cooling at that radius will fall rapidly towards the center. Note that in some of the idealised studies, the fluctuations are seeded by turbulence from AGN \citep[e.g.,][]{Prasad2017,Prasad2018, Prasad2020}. In a cosmological cluster, such as ours, the fluctuations are generated by mergers, movements of galaxies within the ICM, and penetrating streams from the cosmic web \citep[e.g.,][]{Nagai2011,Nagai2013,Zhuravleva2013,Zinger2016,Lau2017}, in addition to feedback. 

Figure~\ref{fig:multiphase} shows the distribution of entropy of all the gas particles within 500 kpc of the halo centre during each of the four epochs. Each cell in this entropy-radius phase space is colored by the log of the probability density function, which is the gas mass in the cell divided by the total gas mass within 500~kpc. The mass-weighted profile of the hot ($T>10^6$K) X-ray emitting gas is plotted in purple. Particles that were close to the central SMBH at any recorded snapshot are shown in red. These particles are a subset of those that have received direct thermal energy injection from AGN feedback, since intermediate \textbf{timesteps} are not recorded. \textbf{We treat these particles instead as tracers of gas that have received feedback more or less directly from the SMBH and these results show that such gas is quickly evacuated} to larger radii. 

The upper two panels in Figure~\ref{fig:multiphase} show a significant amount of low-entropy gas within the central 10s of kpc. Although the AGN is active during these epochs, much of the energy imparted by AGN feedback escapes the core of the cluster via collimated outflows \citep{Tremmel2019}, which is why dramatic changes in AGN activity do not affect the central entropy profiles in these isolated phases. By the third (lower-right) panel, as low-entropy gas is gradually extinguished from the core, the BCG is quenched. \textbf{At this epoch, the central entropy profile increases monotonically with radius, and the dispersion in the central entropy is significantly reduced. Interestingly, this development corresponds to quenching of star formation in the central galaxy. }
%\textbf{An explosive feedback event is not needed to quench the BCG. Instead, this is done gently as star formation uses up the cooler gas already in the BCG, while feedback from the AGN regulates prevents overcooling. The mass-weighted average entropy of the hot phase remains similar to the previous epochs. But, the entropy profile in the central 10kpc has a positive gradient. Following the above results from idealised simulations, this should lead to a decrease in 'precipitation' of low-entropy gas. That is what we see in our cosmological simulation, too}. 
In the final panel, we see the effect of the merger, which unlike AGN feedback raises the entropy of the X-ray emitting hot phase in the central 100 kpc and eliminates the cold phase entirely.  We see in Figure~\ref{fig:bcg_history} that during this phase the central SMBH accretion also decreases by more than an order of magnitude, so it cannot be AGN activity that heated the core.

\begin{figure*}
\begin{center}
\includegraphics[width =\textwidth]{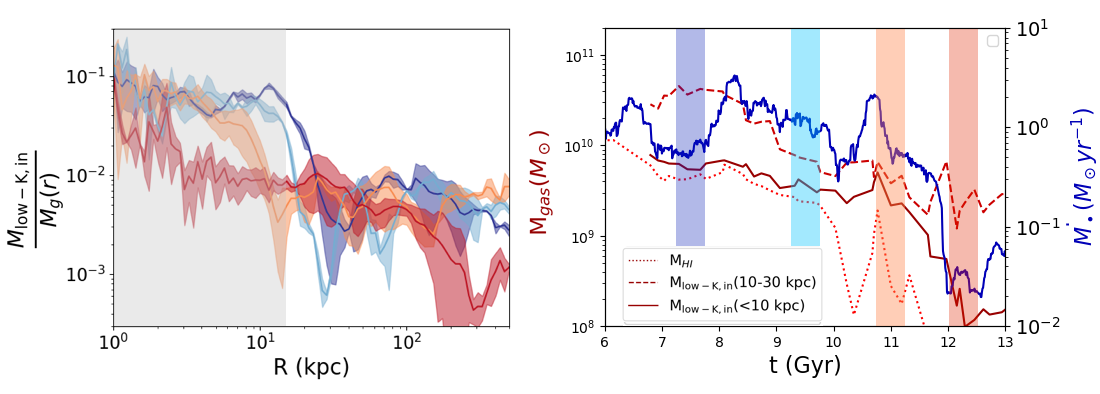}
\caption{
\textit{Left panel:} Mass of infalling, cooling gas, normalised by the total gas mass, as a function of radius during the four epochs. The cooling gas is defined as that with entropy one standard deviation less than the mass-weighted average at a given radius; alternate definitions of cooling gas yielded the same qualitative result. Normalising by the total gas mass accounts for changes in the core gas density, seen in Figure~\ref{fig:ICMprofiles}. The cooling gas fraction is about ten times higher in the central 10-15 kpc before the merger than after; the cooling gas mass fraction mononically decreases with radius, confirming that there is less cooling in the outskirts than in the core. \textit{Right panel:} The total amount of infalling cooling gas, i.e., the left panel integrated within the central 10~kpc (solid) and in 10-30kpc (dashed) as a function of time. The dotted line shows the mass of neutral hydrogen (HI, \textbf{which is explicitly modelled in the simulation}) within the central 10kpc, to show that \textbf{this lower-entropy} gas does correlate to our definition of cooling gas. Black hole accretion rate is plotted on the right axis as in Figure~\ref{fig:core_evolution}. Consisent with results presented in \citet{Tremmel2019}, during the period of AGN activity from $\sim8-11$ Gyr the amount of \textbf{lower-entropy gas, as well as neutral hydrogen,} in the central regions of the cluster declines, eventually leading to the quenching of star formation while the AGN remains active. The merger event represents a more complete disruption of cooling and extinguishes the AGN activities.} 
\label{fig:precip_profile}
\end{center}
\end{figure*}

The difference between panels 3 and 4 is crucial. In the former, the BCG is quenched, the core is running out of cold-phase gas, but the entropy profile of the hot-phase gas is still relatively low in the center. The latter is after the shock heating at 11.97 Gyr, witnessed in Figure~\ref{fig:shock}. There is now no gas with entropy lower than 10 $\rm keV cm^{2}$ within the central 50 kpc, and the entropy of the hot phase is entirely flat in the central 150~kpc. These are very different dynamical states of the cluster.

\textbf{Observations and idealised simulations suggest that the gas cooling is regulated by the AGN-ICM feedback loop. Multi-phase structure in the ICM arises naturally when the heating and cooling balance is globally maintained when the ratio $t_{\rm cool}/t_{\rm ff}$ falls in the range $\sim 5-30$ \citep{McCourt2012,Sharma2012,Prasad2015,Voit2015, Hogan2017, Pulido2018}. Under such conditions, local density enhancements driven, for example, by turbulent  velocity perturbations \citep[c.f.,][and references therein]{Gaspari2013,Prasad2017,Prasad2018} are thermally unstable and condense into clouds that then decouple from their surroundings and 'rain' down upon the BCG, driving star formation and powering the central AGN \citep{Gaspari2013, Prasad2015, Prasad2017, Prasad2018, Li2015, Tremblay2016}.} 
%The reasoning is that if the ratio is sufficiently low, by the time the gas falls to some smaller radius with a lower ambient entropy, it has already cooled to an even lower entropy, and the infall becomes a runaway process all the way into the core.} 

We find that this ratio is indeed important in {\sc RomulusC}. The top panel of Figure~\ref{fig:tcool_tff} shows the ratio of cooling time ($t_{\rm cool}$) to free-fall time ($t_{\rm ff} = \sqrt{2r/g}$). In the first epoch, when AGN feedback is relatively quiet and the BCG is still star forming, the ratio is significantly lower, falling to values as low as $\sim10$. The subsequent steps, during which the BCG is in the process of quenching and the AGN is active, the ratio hovers around $\sim30$. This is consistent with the studies above, which see H-$\alpha$ emitting 10$^4$K gas, indicative of active cooling \citep[][]{Sanders2008,McDonald2010, Fabian2011}, present primarily in groups and clusters with $t_{\rm cool}/t_{\rm ff} \approx 10-30$, marked with the dotted lines in Figure~\ref{fig:tcool_tff}. We acknowledge that our cooling times are overestimated since we do not include metal line cooling in the simulation. We discuss this caveat further in \S4.4. However, the prediction that the transition from star forming to quenching is coincident with a change in the ratio of cooling and free-fall times should be robust to these details. 

The bottom panel of Figure~\ref{fig:tcool_tff} compares the cooling time to the Hubble time, t$_H$. Prior to the merger, even when the BCG is actively quenching (light blue and orange profiles), the cooling time is sub-Gyr. The $t_{\rm cool}$ follows a power law profile at all pre-merger epochs, and central cooling time is less than 1~Gyr. Even after the merger, the cooling time is still of order 10$\%$ of the Hubble time. 
%\textbf{In the absence of heating mechanisms, a low-entropy core would thus reform in such a short time; but the merger, with its shocks and sloshing motions, is still ongoing.} 
The simulation ends before the merger finishes, and it is likely that the continual heating of gas through shocks and turbulent dissipation through sloshing in the core would lead to the re-formation of the original ICM structure over several Gyrs \citep{Poole2006, Poole2008}.

\begin{figure}
\includegraphics[width=0.475\textwidth]{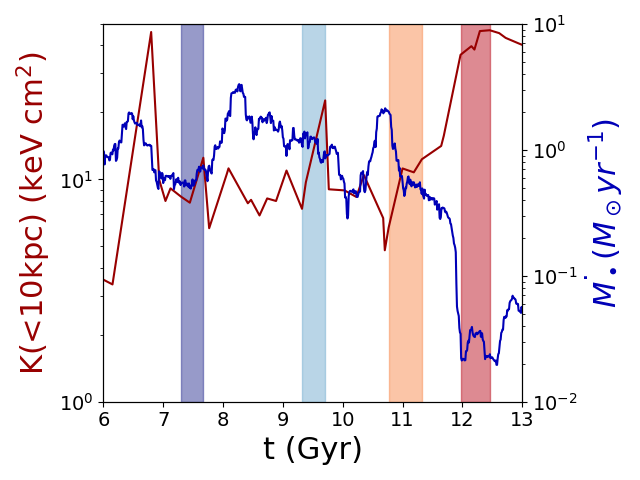}
\caption{Mass-weighted average entropy within the central 10kpc is plotted in red on the left axis, while the black hole accretion rate is plotted in blue on the right axis. The merger at $\sim12$ Gyr results in a rapid change in entropy and SMBH activity.}
\label{fig:core_evolution}
\end{figure}

In order to capture the evolution of gas cooling out of the hot ICM in {\sc RomulusC}, we define gas as "potentially cooling" if it (i) has lower entropy compared to the mass-weighted mean entropy of gas at its cluster-centric radius, i.e., $K < \left[\bar{K}(r) - \sigma_K(r)\right]$, and (ii) is radially in-falling toward the cluster center, $v_r < 0$. Radial bins with small numbers of gas particles, typically those at the center of the cluster, can have large values for the standard deviation of entropy, $\sigma_K$. When this occurs we consider all in-falling gas with entropy $K < \bar{K}(r)/2$. While not a precise value, as it is difficult to determine {\it a priori} which gas particles will effectively cool, this provides a rough baseline with which to compare the ICM evolution. In Figure~\ref{fig:precip_profile} we show this profile of in-falling, relatively low entropy gas as a fraction of the total gas at the corresponding radius. This quantity does not change outside 20 kpc at all epochs, consistent with the unchanging entropy and density profiles at scales above $r\gtrsim 0.2 R_{500}$ as shown in Figure~\ref{fig:ICMprofiles}. At these scales we are only capturing random gas motions and the natural scatter in entropy. However, within $\sim$10-20~kpc depending on the epoch, while the cluster is isolated, the fraction of cooling gas jumps an order of magnitude higher than in the outskirts. Here in the core the ICM, \textbf{gas is more susceptible to cooling based on the entropy and $t_{\rm cool}/t_{\rm ff}$ criteria discussed above}. This region of enhanced cooling decreases from the star forming, through quenching (light blue) and quenched (orange) epochs. After the merger, however, it disappears entirely. The cooling gas fraction within 10kpc is over five times lower post-merger than in the epoch immediately prior.

On the right-hand panel of Figure~\ref{fig:precip_profile}, we show the total mass of such cooling gas within the central 10 kpc (solid red),  $r=(10-30)$ kpc (dashed red) and HI (dotted red). As the cluster goes from star forming to quenching and then quenched, there is less and less cooler gas (HI, molecular hydrogen, etc.) available in the core for star formation. In the inner 50 kpc of the cluster between $t=8$~Gyr and $t=10$~Gyr, where the majority of the quenching takes place, the "potentially cooling" gas mass decreases by 10$^{10.3} M_\odot$ and stellar mass increases by 10$^{10.7} M_\odot$. The increased stellar mass more than accounts \textbf{for this loss of low-entropy gas.} \citet{Tremmel2019} discusses in detail the decrease in both sSFR and cooler gas mass around this time.  Meanwhile, the accretion rate of the black hole (blue solid) slowly declines as well. Once the merger occurs, accretion onto the black hole, \textbf{which in practice is fueled by this low-entropy gas in the center,} is extinguished. This is in line with observations that find that AGN activity is much more common in clusters with low rather than high central entropy \citep{Edwards2007, Sun2009, Mittal2009, Hlavacek-Larrondo2012}.

\subsection{Disruption of the ICM structure}
Early baryonic simulations suffered from an overcooling problem, where too much cold gas accumulated in the cores of galaxies and galaxy clusters (e.g.,~\citealt{Lewis2000, Nagai2007, Borgani2008, Vazza2011a, Borgani2011}). Since then, several phenomenological models of AGN feedback have succeeded in producing realistic gas and star formation profiles and galaxy luminosity distributions in cosmological simulations \citep{Brighenti2006, Sijacki2007, Vazza2013, Weinberger2017}, and some are even able to produce fractions of CC/NCC clusters that are in agreement with the observations \citep{Rasia2015, Hahn2017, Barnes2018}. Still, a clear sense of how this distribution is established and maintained remains elusive. The high resolution of {\sc RomulusC} allows us to understand the mechanism of AGN feedback by directly following the outflows launched by it and studying their impacts on the surrounding gas.

\begin{figure}
    \includegraphics[width=0.475\textwidth]{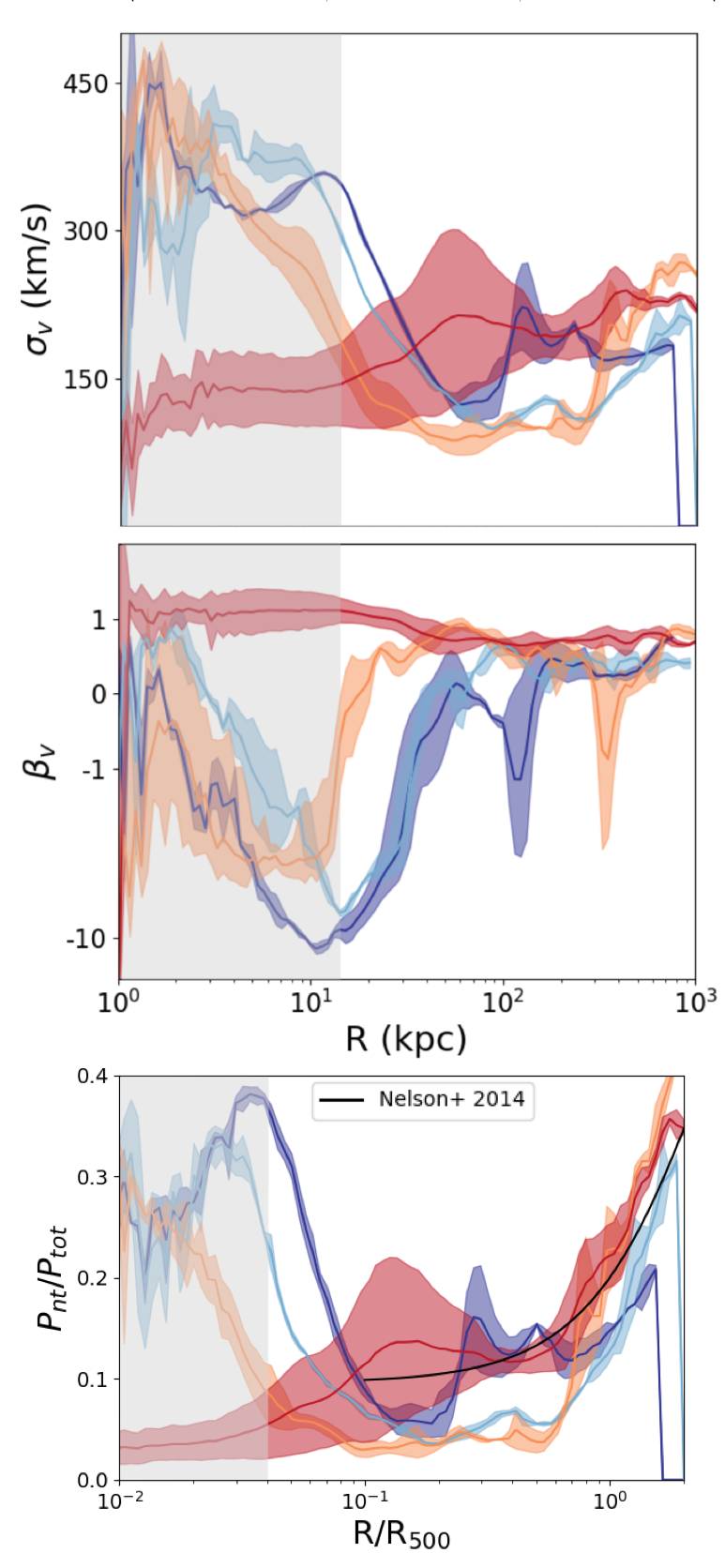}
    \caption{{\it Top panel:}  The average one-dimensional velocity dispersion as a function of cluster-centric distance.   \textit{Middle panel}:  The anisotropy parameter $\beta$ for the motions in the ICM; $\beta$ = 1 for perfectly radial motions, and $\beta = -\infty$ for purely tangential motions. Before the epoch of merger, we now see clearly that the high velocity dispersion in the central 10kpc originates from rotational motions, associated with the BCG gas disk, which lead to a very negative $\beta$. 
    {\it Bottom panel:}  The fraction of non-thermal pressure support for cluster gas compared to the results from Omega500 zoom-in hydrodynamical cosmological simulations of galaxy clusters \citep{Nelson2014}. Rotational motion and AGN outflows result in strong non-thermal support within $r\lesssim 0.1R_{500}$. The merger event disrupts this non-thermal support in the core. Outside $r>0.1R_{500}$, the non-thermal pressure fraction increases monotonically toward cluster outskirts.} 
\label{fig:Pnt}
\end{figure}

We find that a period of enhanced AGN activity is associated with the quenching of star formation (as shown in Figure~\ref{fig:bcg_history}) and the lack of cooler gas in the cluster core (as shown in Figure~\ref{fig:precip_profile}). Even in this dynamic environment, however, Figure~\ref{fig:mergermaps} shows that the entropy within $\sim$10kpc remains below 10 $\rm keV~cm^2$ until the merger. One reason, as shown in Figure~\ref{fig:multiphase}, is that particles receiving thermal energy from the AGN quickly move out to large radii until they are in equilibrium with their surroundings, instead of dissipating that heat to other gas particles in the core. Furthermore, the energy imparted by the AGN over characteristic cooling times of 10 Myr is $\sim 10^{61}$ ergs, a mere 0.1$\%$ of the $10^{64}$ergs required to heat all the gas within the central 0.1$R_{500}$ to the post-merger entropy of 100 keV cm$^2$. We note that the AGN at the core of RomulusC is not particularly weak, and at its brightest is comparable to most luminous low-redshift AGN known, which lives in the galaxy cluster MS-0735 \citep{Gitti2007}.

Galaxy cluster mergers generate significant bulk and turbulent gas motions, which in turn can be converted into heating of the ICM through a combination of mixing and turbulent dissipation \citep{Poole2006, Poole2008, ZuHone2010, ZuHone2011}. \textbf{The top panel of figure~\ref{fig:Pnt} shows the one-dimensional velocity dispersion of all the gas in {\sc RomulusC}, assuming an isotropic velocity distribution ($\sigma_v = \sigma_{3D}/\sqrt{3}$)}. The velocity dispersion decreases significantly in the central $\sim$20~kpc after the merger, and increases outside of this region. Prior to the merger, this central region is rotationally supported, with large tangential velocities associated with the rotating BCG gas disk. After the merger, this rotational support is disrupted, as are the powerful radial outflows generated by the central SMBH, which also becomes significantly quieter during this epoch.

We verify this explanation of the evolution of the gas dynamics in the core by breaking up the motions into tangential and radial components. The middle panel shows the velocity anisotropy parameter $\beta \equiv 1- 0.5 \sigma_t^2 / \sigma_r^2$, where $\sigma_t$ and $\sigma_r$ are the radial and tangential velocity dispersions, respectively. Note that $\beta=1$ for purely radial motions, and $\beta \rightarrow -\infty$ for purely tangential motions. Outside the central 0.1$R_{500}$, before the merger, the velocity dispersion flattens around 100 km/s, consistent with gas motions generated by mergers and movement of galaxy motions through the ICM \citep[e.g.,][]{Ruszkowski2011,Nagai2013}. \textbf{The bump around 100~kpc for the earliest (dark blue) epoch is due to rotational motions.} The one-dimensional average velocity dispersion in the central 0.1$R_{500}$ of {\sc RomulusC} is 200-400 km/s. Breaking this down into radial and tangential components, we find that the higher values in the center are dominated by the rotation of the BCG. Outside the core, and at all radii after the merger, the values are consistent with the 150-200 km/s measured by Hitomi in the Perseus cluster \citep{Hitomi2018} and numerical simulations \citep{Lau2017,Bourne2017}. 

Bulk motions of the gas are also a source of pressure support against the gravitational potential of the cluster, in addition to the thermal pressure of the ICM. This non-thermal pressure fraction, $P_{\rm nt} / P_{\rm tot} = \mu m_p\sigma_{v}^2 / (k_BT + m_p\sigma_{v}^2)$, translates into the deviation from hydrostatic equilibrium, which is a key assumption in mass measurements both from X-ray and SZ observations \citep[e.g.,][]{Nagai2007b,Mahdavi2008,Mahdavi2013,Hoekstra2015,Biffi2016,Shi2016,Pearce2020,Angelinelli2020}. Large-box cosmological simulations to date have found that this non-thermal pressure fraction increases with halo mass, and increases with cluster-centric distance, since mergers and accretion are a major source of this pressure are more frequent in high mass clusters \citep{Lau2009,Nelson2014,Shi2015}. Consistent with these results, \citet{Mahdavi2013} find via a joint X-ray and weak lensing analysis of 50 observed galaxy cluster systems that while CC clusters show no evidence of hydrostatic mass bias, NCC clusters exhibit a 15\%-20\% bias between $R_{2500}$ and $R_{500}$.

The non-thermal pressure fraction for {\sc RomulusC} is shown as a function of the cluster-centric radius in the bottom panel of Figure~\ref{fig:Pnt}. For comparison, the black line shows the average non-thermal pressure fraction of 65 massive galaxy clusters from the {\em Omega500} non-radiative simulation \citep{Nelson2014}. The merger does increase the non-thermal pressure support outside 0.1$R_{500}$, but within the core ($r\lesssim 0.1R_{500}$) this is actually reduced, since, as described above, the central region was dominated by strong tangential gas motions associated with gaseous disk of the BCG and outflows from the AGN, both of which are disrupted by the merger.

\section{Discussion}
\label{sec:discussion}

\subsection{Comparison with Other Cosmological Simulations}

Cosmological simulations of galaxy clusters provide a realistic environment and assembly history within which to study their evolution, self consistently modeling the effects of both feedback processes, cosmological accretion and sub-structures on the ICM, and BCG evolution. Generally, the major limiting factor of such simulation is the lack of sufficient resolution to resolve all relevant dynamic ranges of this multi-scale, multi-physics problem. A major strength of {\sc RomulusC} is its resolution, matched only by the TNG50 simulation \citep{Nelson2019} to date. This is critical for a proper treatment of both feedback processes, as well as to \textbf{reproduce the distribution of cooler ($T < 10^5K$) gas} in the ICM \citep{Butsky2019}.

The TNG suite implements AGN feedback as an isotropic momentum kick instead of a thermal injection with a local cooling shutoff; in a comparative study of different sub-grid heating models \citep{Smith2017}, these two prescriptions were found to have the best convergence with resolution. The differences due to both the hydrodynamic solver and feedback implementation, however, are expected to decrease at higher resolutions. Indeed, similar to {\sc RomulusC}, \citet{Nelson2019} find that even though the energy input is isotropic, the outflows are collimated as they follow the path of least resistance perpendicular to the gas disk of the galaxy. They also find that the AGN activity quenches star formation in the late stages of the galaxy's evolution, but continues after star formation is quenched. 
In the larger box of Illustris-TNG300, \citet{Barnes2018} found that CCs are not more relaxed than NCCs, suggesting that mergers may not be solely responsible for disrupting CCs. This is in contrast to recent works \citep[e.g.,]{Rasia2015, Hahn2017} that produce realistic CC/NCC fractions and agree with our finding that mergers play a crucial role in increasing and flattening the entropy of cluster cores. 
This is an interesting discrepancy that should be explored in further work. 

\subsection{Relation to Theoretical Models for AGN-regulated Cooling}

The evolution of cooling to free-fall time for {\sc RomulusC} presented in Figure~\ref{fig:tcool_tff} shows that our results are broadly consistent with the picture of precipitation-regulated cooling \citep[c.f.,][]{Sharma2012, Gaspari2013,Gaspari2017, Voit2015, Prasad2017, Prasad2018}. Typical cooling to free-fall time ratios for observed clusters lie in a range $\sim10-30$ \citep{Hogan2017,Voit2018b}, though values below 10 have been observed \citep{Pulido2018, Babyk2018}. Ratios below 30 are seen in clusters with significant multiphase gas, whereas values around 30 or above are coincident with ICM that are mostly single phase \citep{Voit2018b}. As shown in Figure~\ref{fig:tcool_tff}, the $t_{\rm cool}/t_{\rm ff}$ ratios in the {\sc RomulusC} are typically no smaller than 10 and, but the ratios fall well below 30 during the phase with significant cooling and star formation in the core. During the period when the BCG is quenching due to strong, large-scale outflows driven by the AGN \citep{Tremmel2019} the ratio remains steady at or just below $\sim30$, consistent with observed clusters with less multiphase gas in their centers.

As shown in \citet{Butsky2019}, the ICM of {\sc RomulusC} contains \textbf{ both the X-ray emitting ($T > 10^6K$) and the cooler ($T<10^5K$) phases}.
%, which is why it is important to have such high resolution in order to fully resolve gas cooling. 
However, despite the state-of-the-art resolution, the current simulation still lacks the resolution to fully resolve the thermal instabilities. Despite these uncertainties, the fact that we see AGN-regulated gas cooling that quenches star formation with realistic values of $t_{\rm cool}/t_{\rm ff}$ in a fully cosmological simulation is an important result and further supports this theoretical picture of CC clusters.

\subsection{Thermodynamics of the Cluster Core}
One of our main conclusions is that AGN activity alone is not capable of significantly increasing the entropy within the entire core region of {\sc RomulusC}, or completely shutting off cooling. This is in contrast to the results of idealized simulations of AGN feedback in the ICM \citep[e.g.,][]{Li2015, li_ruszkowski_bryan_2016}, where gas in the core goes through cycles of over- and under-heating by the AGN, corresponding to CC and NCC states of the core gas. There can be two factors at play here. First, these idealised simulations did not have turbulence other than that sourced from the AGN outflows. \citet{Gaspari2013} and \citet{Prasad2017} have shown that the presence of turbulence in the ICM enhances the formation of cold phase gas by 1-2 orders of magnitude. In the absence of this turbulence seeding additional thermal instability at larger radii, such idealised simulations have a much easier time overheating the core, since the core is not replenished as efficiently by inflows. Secondly, few idealised simulations have included a BCG, such as the one that forms in {\sc RomulusC}. \citet{Prasad2018} found that adding a BCG potential increases the core density and decreases the core entropy, fortifying a CC structure and making it harder to disrupt. 

The merger fundamentally disrupts the AGN-ICM equilibrium in at least three key ways. First, it shocks the gas in the core \citep{Poole2006, McCarthy2007}. We see this as a overpressured core in Figure~\ref{fig:shock}, which expands and dissipates by the following snapshot, when the ICM structure is clearly disrupted. Secondly, the impact of the merging event physically removes cooler gas from the core of the main cluster and mixes high-entropy gas from larger radii with low-entropy gas in the core through gas sloshing \citep{Poole2006,Poole2008,ZuHone2010,ZuHone2011}.  
Third, this sloshing cascades into smaller scale turbulent motions which in turn dissipate into heat \citep{Miniati2014, Banerjee2014,Wittor2017}. It is difficult to assess the relative importance of these processes in a cosmological simulation like {\sc RomulusC}, because the effects of mergers as well as feedback are highly intertwined. 
Future work should focus on quantifying the role of each heating channel and its detailed mechanism using idealized simulations, in which each process is added and investigated systematically one at a time.

\subsection{Gas Motions in the Cluster Core}

The gas velocities in the center of {\sc RomulusC} before the merger are larger that those observed in the Perseus cluster using Hitomi \citep{Hitomi2018}, though Hitomi's spectroscopic measurement of the Perseus alone does not strongly constrain the rotational motions in cluster cores. While large gaseous disks have been observed in the centers of massive halos \citep[e.g.][]{Hamer2014, Nagai2019}, they are often difficult to detect. Observations of molecular and neutral gas in the center of CC clusters have been observed to have significant rotational velocities \citep{Russell2019}. As the AGN feedback regulates gas cooling on larger scales in {\sc RomulusC}, the rotationally supported core shrinks by a factor of $\sim2$, evident from the evolution shown in Figure~\ref{fig:Pnt}.

So far, a rotating disk of cold gas has been \textbf{resolved in the BCG of the Perseus cluster \citep{Nagai2019} and Hydra A \citep{Hamer2014}; \citet{Hamer2016} further find kinematic evidence for central rotating disks of cold gas in 2/3 of the galaxy clusters and groups they observed.} These disks are extremely difficult to see using only X-ray telescopes, because the hotter gas dominates the luminosity, and current telescopes lack the dynamic range to resolve both the disk and surrounding ICM; furthermore, the response function of current X-ray telescopes falls off sharply below $1$keV, and a disk in equilibrium with the BCG potential would be much cooler than this. Such X-ray only observations may be possible with future instruments, such as Lynx \citep{Gaskin2019}, which, in addition to a strong response at low temperatures and a significantly greater dynamic range, would capture X-ray spectra and allow the rotating component to be separated from the quasi-static ICM. 

The model for AGN feedback may also affect this structure and the presence of such a rotationally supported disk may be explained by feedback that is too inefficient, potentially supporting the need for variable efficiency similar to two-mode models. Important for our results, the presence of such rotational support likely helps keep the core stable against disruption from the AGN, requiring instead a more catastrophic event like a head-on merger. Differences from other idealized and cosmological simulations already point to a dependence on AGN feedback prescription for the evolution of cluster cores. However, we stress that the result that AGN feedback regulates cooling and star formation while maintaining the cluster's core structure remains important and robust despite these uncertainties. Understanding the effects of different AGN implementations will require further comparison with results from cosmological simulations of similar resolution (e.g., TNG50) as well as a larger sample of high resolution simulations of massive halos.

Finally, RomulusC cannot show that the core heating is permanent, since the simulation ends 2 Gyr after the merger begins. At $r=0.1R_{500}$ in Figure~\ref{fig:tcool_tff}, the cooling time is still $\sim$ 3 Gyr, so that to check for a permanent transformation, we would have to run the simulation forward another 2-3 Gyr. What we can say is that this one, relative low mass ratio, head-on merger disrupts the cluster core for at least 3 Gyr.

\subsection{Caveats \& Future Work}

As discussed in \citet{Tremmel2019}, and briefly here in \S2.1, the {\sc Romulus} simulations do not include the effects of high temperature metal lines, an important coolant in the hot gas of massive halos. This will primarily affect the accretion history of gas onto the central galaxy, though this process is affected by the details of stellar and AGN feedback implementations as well \citep{vdVoort2011a, vdVoort2011b}. Were we to have included metal line cooling, gas feeding star formation and SMBH activity at later times might have cooled sooner which would affect when quenching takes place as well as the physical state of the ICM at the time of the merger. However, we stress that even without metal line cooling, the cooling times in the core of {\sc RomulusC} prior to the merger are sub-Gyr (Figure
~\ref{fig:tcool_tff}), meaning that this alone cannot explain the regulation of cooling and that additional heating from both AGN and the merger event is required to regulate and quench star formation. Still, the presence of additional cooling processes may affect the detailed evolution of the $t_{\rm cool}/t_{\rm ff}$ ratio. The relative evolution of this ratio within the framework of this simulation and its coincidence with the onset of quenching remains an important prediction in agreement with both theoretical models and observations. It is likely that different cooling models, as with different feedback recipes, will result in different star formation and SMBH growth histories, as well as different $t_{\rm cool}/t_{\rm ff}$ ratios. Higher resolution simulations capable of resolving molecular gas and metal line cooling are still needed to make more accurate predictions, though such simulations will still be sensitive to {\it ad hoc} choices in AGN feedback prescriptions. While more advanced simulations are needed to make more detailed predictions, the fact that AGN feedback is capable of regulating gas cooling on large scales without disrupting the ICM core structure is an important proof of concept, along with the connection between quenching and $t_{\rm cool}/t_{\rm ff}$.

The details of AGN feedback are highly uncertain given that all of the relevant microphysics are unresolved at even the high resolution of {\sc RomulusC}. The choice to include only thermal feedback with a cooling shutoff (see \S2.2.2) means that the large-scale kinetic feedback that we observe in the simulation is driven by hot, buoyant gas, which may be different if kinetic feedback were implemented directly. Simulations which implement a `two-mode' feedback prescription \citep[e.g.,][]{Weinberger2017} will generally change the feedback efficiency for black holes of different mass. In {\sc RomulusC} our black hole feedback prescription assumes a constant proportionality with accretion rate, meaning that high accretion rates are required for strong feedback to occur. Since this efficiency was optimized to produce broadly realistic galaxies \citep{Tremmel2017}, changing AGN feedback efficiency within uncertainties (i.e., without significantly disturbing the galaxy properties) will likely alter the detailed star formation and SMBH growth histories of the BCG. However, our results regarding AGN feedback and its ability to regulate large-scale cooling should be insensitive to the details of the model, as our sub-grid physics affects gas at 100s pc scales while our results pertain to 10s-100 kpc scales in the simulation.

Further, {\sc RomulusC} does not include the effects of magnetic fields or cosmic rays, which have been shown to play an important role in transfering feedback energy into the ICM. \citet{Ruszkowski2010}, for example, showed that in the presence of tangled magnetic fields, as expected in a turbulent ICM, thermal conduction from the hot outskirts to the core can be significant. \citet{Enslin2011} showed similarly that their simulated CC structure was weakened by cosmic rays. Each of these would make it easier to form a high, flat-entropy core using AGN activity alone, since the core would never get as cold and dense as in their absence \citep[e.g.,][for a review]{Nulsen1982,Sarazin2004, mcnamara_nulsen_2007}. The role of each of these physical processes is currently being explored by idealised simulations \citep[e.g.,][]{Ruszkowski2007,Sharma2009,Parrish2009,Enslin2011,Parrish2012,ZuHone2013, Kannan2016, Yang2016b}. Further testing of the effects of different feedback mechanisms will be important. The results of this paper should be seen as a proof-of-concept that it is possible to regulate cooling and star formation with large-scale AGN feedback without disrupting the dense core structure in the ICM.

\section{Conclusions}
\label{sec:conclusions}

Using the {\sc RomulusC} high resolution cosmological simulation of a $10^{14}$ M$_{\odot}$ galaxy cluster we study the relative roles of AGN feedback and major mergers in regulating star formation and determining the structure of the ICM. With its unprecedented resolution, {\sc RomulusC} produces gas particles \textbf{as cool as $10^4K$} \citep{Butsky2019} and naturally produce large-scale collimated outflows from AGN that are able to quench star formation without significantly changing the structure of the ICM \citep{Tremmel2019}. The simulation also undergoes a merger event at $z\sim0.14$ that is still on-going at $z=0$, which, in contrast with AGN feedback, results in a significant change to the ICM structure and cooling efficiency. In this work we focus on understanding in more detail how AGN feedback is able to regulate star formation in the BCG while co-existing with a low entropy ICM core and contrast this with the effect of the merger event. In order to do this, we examine the simulation within four time bins representing different phases of evolution: 1) BCG is star forming, 2) BCG is actively quenching, 3) the BCG is fully quenched, and 4) onset of the merger event. We find that:

\begin{enumerate}
\item AGN feedback is able to quench star formation by reducing gas cooling at $\sim$ 10-30 kpc scales and steepen the entropy profile while co-existing with a low entropy cluster core and without disrupting the structure of the ICM.
\item A ratio $t_{\rm cool}/t_{\rm ff} \sim$ 30 represents a critical transition between a rapidly cooling cluster core with a star-forming BCG and more moderate cooling with a quenching/quenched BCG, consistent with both theory and observations. The exact value of this threshold may change upon inclusion of metal line cooling in our model.
\item Gas particles are heated directly by the AGN rise buoyantly to radii $>30$kpc on short timescales, so that their much of their heat is dissipated on larger scales.
\item A head-on, 1:8 merger is highly disruptive to the ICM struture, resulting in a sudden increase and flattening of the core entropy and a marked decline in AGN activity by 1-2 orders of magnitude.
\item Cooling times in the core during the merger are of order 10$\%$ the Hubble time. While energy dissipated from the on-going merger actively prevents this cooling through $z=0$ it is likely that the core will relax again shortly after the completion of the merger.
\end{enumerate}

This work demonstrates that AGN feedback can regulate cooling within cluster cores without disrupting the structure of the inner ICM. This is in contrast with a massive, head-on merger which does significantly disrupt, at least for a few Gyr, both the ICM structure and the regulating loop between AGN and gas cooling.

Prior to its disruption by the merger, the gas within the low entropy core of the simulated cluster showed strong rotational support. Such disks have been detected in 10-20$\%$ of systems via cold molecular gas, and have only been observed in two systems in X-ray; these could provide stability to cluster cores. Future X-ray spectroscopic missions such as Athena and Lynx may reveal the warm-hot component of rotating gaseous disk around BCG. The merger quickly destroys this rotational structure.

Ultimately, {\sc RomulusC} is a simulation of a single cluster, and a complete comparison to observations will require simulations of a larger sample of galaxy clusters. A significantly larger volume with similar resolution is hard to achieve given current computational constraints. A promising alternative lies in the genetic modification (GM) technique of \citet{Roth2016}, which would allow us to systematically vary individual merger parameters such as mass ratio, impact parameter, and infall velocity, in a controlled fashion currently restricted to idealised simulations. 

\section*{Acknowledgements}
We thank Paul Nulsen, Ming Sun, Grant Tremblay and John ZuHone for useful discussions. Analysis was conducted primarily on the NASA Pleiades computer with assistance from Alyson Brooks and using the publicly available tools {\sc Pynbody} \citep{Pontzen2013} and {\sc TANGOS} \citep{Pontzen2018}.

This research is part of the Blue Waters sustained-petascale computing project, which is supported by the National Science Foundation (awards OCI-0725070 and ACI-1238993) and the state of Illinois. Blue Waters is a joint effort of the University of Illinois at Urbana-Champaign and its National Center for Supercomputing Applications. This work is also part of a Petascale Computing Resource Allocations allocation support by the National Science Foundation (award number OAC-1613674).  This work also used the Extreme Science and Engineering Discovery Environment (XSEDE), which is supported by National Science Foundation grant number ACI-1548562. Resources supporting this work were also provided by the NASA High-End Computing (HEC) Program through the NASA Advanced Supercomputing (NAS) Division at Ames Research Center.

UC was supported in part by the Chandra X-ray Center Predoctoral Fellowship. AB,  TQ  and  MT  were  partially  supported  by NSF award AST-1514868, and AB also acknowledge support from  NSERC (Canada) through  the  Discovery  Grant  program. MT gratefully acknowledges the support of the Yale Center for Astronomy and Astrophysics Postdoctoral Fellowship. This work is supported in part by the facilities and staff of the Yale Center for Research Computing.

\section{Data Availability}
The data underlying this article will be shared on reasonable request to the corresponding author.

\bibliographystyle{mnras}
\bibliography{reference}
% Don't change these lines
\bsp	% typesetting comment
\label{lastpage}
\end{document}